# Large Scale Parallelization Using File-Based Communications

Chansup Byun, Jeremy Kepner, William Arcand, David Bestor, Bill Bergeron, Vijay Gadepally, Michael Houle, Matthew Hubbell,
Michael Jones, Anna Klein, Peter Michaleas, Julie Mullen, Andrew Prout, Antonio Rosa, Siddharth Samsi, Charles Yee, Albert Reuther
MIT Lincoln Laboratory
Lexington, MA, USA
{cbyun, kepner, reuther, warcand, david.bestor, bbergeron, vijayg, michael.houle, mhubbell, michael.jones,
anna.klein, pmichaleas, jsm, aprout, antonio.rosa, sid, yee, reuther}@ll.mit.edu

*Abstract*— **In this paper, we present a novel and new file-based communication architecture using the local filesystem for large scale parallelization. This new approach eliminates the issues with filesystem overload and resource contention when using the central filesystem for large parallel jobs. The new approach incurs additional overhead due to inter-node message file transfers when both the sending and receiving processes are not on the same node. However, even with this additional overhead cost, its benefits are far greater for the overall cluster operation in addition to the performance enhancement in message communications for large scale parallel jobs. For example, when running a 2048-process parallel job, it achieved about 34 times better performance with MPI_Bcast() when using the local filesystem. Furthermore, since the security for transferring message files is handled entirely by using the secure copy protocol (scp) and the file system permissions, no additional security measures or ports are required other than those that are typically required on an HPC system.**

*Keywords—large scale, parallelization, file-based communication, scp, filesystem permission, security*

## I. INTRODUCTION

For the distributed memory systems, a number of parallel programming communication libraries such as MPI [1], PVM [2], and BLACS [3] have been developed. Among those, the MPI standard implementation emerged as the de-facto standard, and it has been widely adopted by many researchers, developers, and industry as a whole. Along a similar thread, there have been many attempts to parallelize MATLAB [4] and GNU Octave [5] on distributed memory systems [6, 7, 8], with the majority of the attempts having been implemented with the MPI standard, mostly based on the socket communication layer. Among these attempts is also a file-based MPI communication scheme for parallelizing MATLAB and Octave codes which was developed by the Lincoln Laboratory Supercomputing Center (LLSC – formerly LLGrid) team. While file-based inter-process communications are not common, a number of researchers, such as Chen, et. al. [9], have published works about using XML-based data for agent communications in mobile systems, where mobile agents roam over the network accessing distributed resources and cooperating with other agents or non-agent components during the course of performing certain tasks. However, to the best of our knowledge, there has been little literatures about using file-based messaging communications for high-performance computing applications.

The MatlabMPI toolbox [10] implemented a file-based communication scheme for interprocess communications, which later become a part of the pMatlab [11] toolbox. pMatlab has become one of the most popular parallelization packages for MATLAB/Octave programming at LLSC since the inception of the interactive, on-demand grid computing environment using gridMatlab and pMatlab [12]. The proliferation of pMatlab has been accomplished by providing an on-demand, interactive, easy-to-use, high-performance grid computing environment, the Lincoln Laboratory Grid (LLGrid) system [13, 14]. This system was designed to provide Laboratory staff with an effective way to exploit cluster computing as a solution to the demand for computational power in large-scale algorithm development, data analysis, and simulation tasks. The current file-based communication implementation, MatlabMPI, is a Matlab implementation of a subset of the Message Passing Interface (MPI) standard [1] and allows any MATLAB/Octave program to exploit multiple processors. pMatlab is a parallel programming toolbox, which consists of a library of objects and routines for distributing numerical arrays onto multiple processors and then carrying out parallel computations on these distributed arrays. A typical MATLAB/Octave programmer can use pMatlab to convert a program to a parallel implementation in a few hours and can then run the application on a cluster. The LLGrid system architecture allows a pMatlab program to be run on a remote cluster as simply as it is to run a MATLAB/Octave program on a desktop. The gridMatlab toolbox interfaces with an underlying resource manager/scheduler for three activities: cluster status monitoring (how many processors are available), job launching, and job aborting [14].

This material is based upon work supported by the Assistant Secretary of Defense for Research and Engineering under Air Force Contract No. FA8721-05-C-0002 and/or FA8702-15-D-0001. Any opinions, findings, conclusions or recommendations expressed in this material are those of the author(s) and do not necessarily reflect the views of the Assistant Secretary of Defense for Research and Engineering.

Since the LLGrid system was first built, it has grown significantly in recent years as we have added more computing and storage capacity in order to meet LLSC users' diverse research and development needs. As the system grows, we have also increased the resource limits per user on our various computing environment. Although increasing users' resource limits helps users run much larger jobs, it also has caused undesirable side effects including frequent high loads on the central filesystem when one or more large size jobs is running concurrently, generating large numbers of file accesses. We have worked with and coached many users to help them become aware of this issue and modify their jobs to eliminate or reduce the central filesystem load. However, one of the many sources of heavy central filesystem loads remains the file-based communication messages via the MatlabMPI messaging kernel, which continue to use the central filesystem as shown in Figure 1.

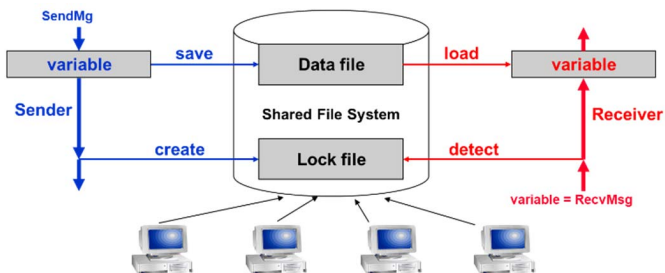

**Figure 1: File-based messaging kernel architecture using a shared central filesystem.**

We have looked at various ways to reduce the central filesystem loads while keeping the current underlying file-based communication architecture. One way to replace the central filesystem as the medium for message transfer, while keeping the file-based messaging kernel is to use a local filesystem. By doing so, we can eliminate most of the heavy file I/O on the central filesystem. We have demonstrated that the file-based messaging kernel performance can be significantly improved by cross-mounting a local filesystem to the node where the receiving process is running [15]. However, this approach is too complex to use on production cluster systems. In the proposed approach, a file transfer utility such as the secure copy protocol (scp) is used when both sending and receiving pMatlab processes are not on the same compute node. There has been an earlier work to incorporate the secure copy utility to transfer files between local Matlab and remote sever by Nehrbass, et. al. [16], but this work is mainly focused on transferring files, not for message communications between processes. It is important to note that, since the security for transferring message files is entirely handled by the scp tool and the file system permissions, no additional security or ports are required other than those that are typically required on an HPC system. In this paper, we will discuss how the new MatlabMPI messaging kernel is implemented using local filesystems and demonstrate its performance compared to the existing MatlabMPI messaging kernel using the central filesystem.

## II. APPROACH

With the current file-based messaging architecture, all parallel processes are writing messages as files to the central filesystem and creating a lock file corresponding to each message file when the data message write has completed as shown in Figure 1. This approach works well with a small cluster system but, as the system grows bigger, the cluster experiences high filesystem loads when a very large parallel job is running. A great deal of the load is the rapid, periodic polling of the many receiving processes (of very large jobs) of the file system to determine whether their lock file has been written. In order to overcome the issue, the current file-based messaging architecture can be modified to use a local filesystem as shown in Figure 2. When the sending and receiving processes are located two different nodes, Node A and Node B, the sending process first creates the message and lock files on its own local filesystem on Node A. Then, it initiates a scp call to transfer the message and lock file (in that order) to Node B. On the receiving side, the receiving process is waiting and polling for the incoming lock file on its own filesystem. As soon as the receiving process detects the lock file on its own local filesystem, it starts reading the message file. In the new file-based messaging architecture, all the message and lock files are written to local filesystems, in a dynamically created directory path stipulated by the scheduler in an environment variable, TMPDIR, when the job gets dispatched to run.

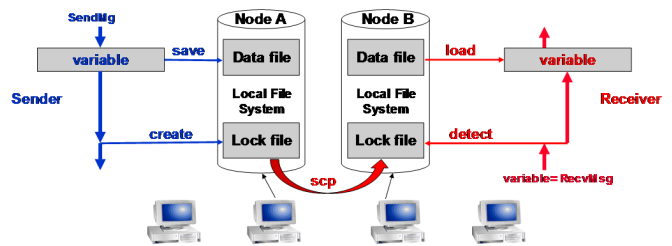

**Figure 2: The new file-based messaging kernel architecture using local filesystems**.

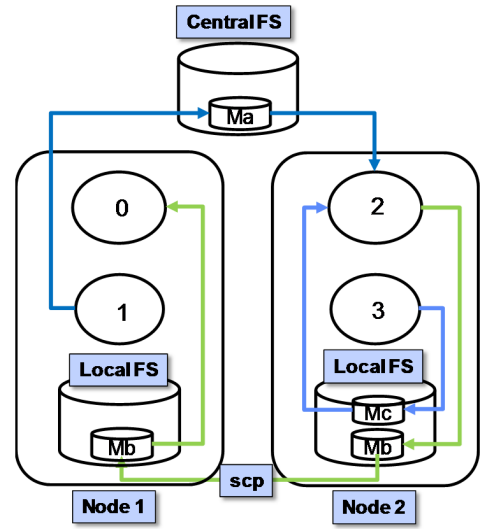

**Figure 3: Send and receive operation comparison when using file-based messaging kernel architecture with the central filesystem and the local filesystem**. **The number inside the circle represents the rank of parallel processes.**

Figure 3 shows three messages (Ma, Mb, and Mc) that are being transferred among four processes (ranks 0, 1, 2, and 3) across two nodes (1 and 2). With the current file-based messaging kernel, there is no knowledge of whether the sending and receiving pMatlab processes are on the same compute node or not or whether both nodes have access to the central filesystem. For message Ma, the sending process (rank 1) writes the message (Ma) into a message file on the central filesystem and then, the receiving process (rank 2) reads the message from the central filesystem, oblivious of what node the message originated it. However, if a local filesystem is used instead of the central filesystem, it is necessary to check whether the sending and receiving parallel processes are on the same node or not. This check is done by creating a host-to-rank map, which contains the information about which compute node each parallel process is running on and the TMPDIR path for each parallel process. Based on the host-to-rank map, if both parallel processes are on the same node such as the parallel processes of ranks 2 and 3 as shown in Figure 3, the sending process (rank 3) writes message Mc as a file on its own local filesystem and the receiving process (rank 2) can simply read the message from the same local filesystem. However, if both processes are not on the same node such as the parallel processes, ranks 0 and 2, the sending process (rank 2) needs to transfer the Mb message to the Node 1, where the receiving process (rank 0) is running, using a file transfer utility, scp, after it writes the Mb message as a file on its own local filesystem first. So if both sending and receiving processes are not on the same compute node, there are additional costs for transferring files. As we will see in a later section, it turns out that this cost is marginal as compared to the performance degradation caused by issues related to high loads on the central filesystem.

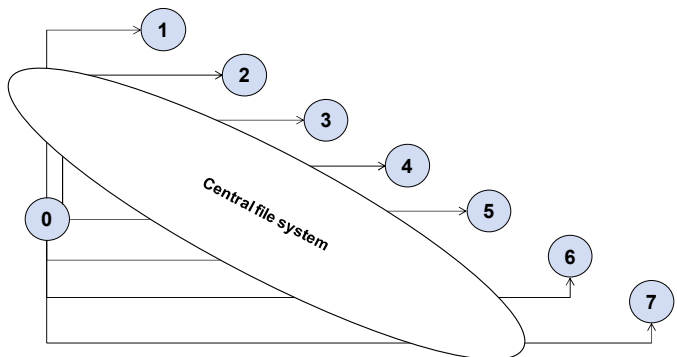

**Figure 4: A schematic diagram for current broadcast operation using a central filesystem**.

The proposed local file-based messaging kernel works well with the point-to-point messages. However, it can have a significant performance impact with the broadcasting operation if its central filesystem is directly replaced with a set of local filesystems. According to the current broadcast operation as shown in Figure 4, the sending process (rank 0) writes a message file on the central filesystem first and then writes a symbolic link to the message file for each of the receiving processes. When that is complete, it does the same thing for lock files. When the lock files have been written, all the receiving processes read the lock file and receive the broadcast message by reading the message (through the symbolic link) from the central filesystem. If the central filesystem is replaced with a set of local filesystems and a simple broadcast is used, the sending process (rank 0) now would need to transfer the message file and lock file to each of the receiving processes. These additional file transfers becomes a serializing bottleneck in the broadcasting operation.

This bottleneck issue can be alleviated by introducing a new broadcast scheme, a so-called node-aware broadcasting algorithm. In this node-aware broadcasting scheme, the leader process of each compute node is identified and then, the broadcasting operation is performed in two levels: first a broadcast among the leader processes and then, a broadcast among the processes within the same node. The leader process is defined as the parallel process with the lowest rank among those processes on the same compute node.

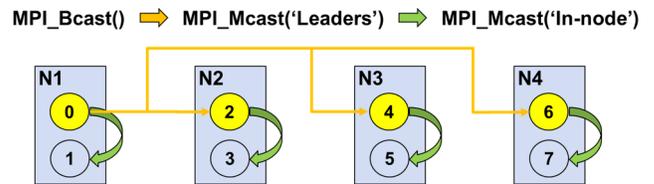

**Figure 5: A schematic diagram for a broadcast operation using a node-aware broadcasting scheme with two-level multi-cast operations.**

Figure 5 shows a node-aware broadcast scheme, which is implemented as two levels of multi-cast operations, one multi-cast operation among the leaders of each compute node and the other multi-cast operation among the parallel processes within the same node. In this example, there are 4 compute nodes (N1 through N4) and each compute node has two parallel processes. The leader processes are the parallel processes of ranks 0, 2, 4 and 6, based on the definition. If the source is the rank 0 process, the first multi-cast operation (writing message and lock files) will be done among the leader processes, where the rank 2, 4, and 6 processes will receive the broadcast message from the rank 0 process. Then, each leader process (ranks 0, 2, 4, and 6) will multi-cast the message to other processes on the same node. In this example, the rank 1 process will receive the broadcast message from the rank 0 process, while rank 3 process will receive the broadcast message from the rank 2 process, and so on. Since the second level multi-cast operation is done within a node, the receiving processes can receive the message by reading the message file from the same local filesystem.

As we will see in a later section, the node-aware broadcasting algorithm can reduce the broadcasting time significantly by reducing the bottleneck associated with the file transfer requirement. In addition, the node-aware broadcasting scheme can reduce the time further (compared to the original scheme using the central filesystem) since the new scheme eliminates resource contention on the central file system when the messages are broadcast to a large number of parallel processes.

Another common collective operation is aggregation. The agg() function aggregates a distributed global array into a double array on the rank 0 process. The current agg() implementation uses a hierarchical binary collection of the distributed global array, as shown in Figure 6, which generates a lot of file I/O as the parallel job size increases. With the current agg() function, the distributed global array is aggregated with a maximum number of message communication of $\log_2 N_p$, where $N_p$ is the total number of parallel processes. Since the underlying communication for the agg() function is point-to-point message communication, the agg() function can be switched to use local filesystems without any modification as soon as the point-to-point communication is changed to use local filesystems. The agg() function behaves well with large parallel jobs when using the local filesystems because it distributes the file I/O to the local filesystems thereby eliminating load on the central filesystem. However, since the current agg() implementation does not know how the parallel processes are distributed across multiple compute nodes, it may cause unnecessary remote file transfer calls unless the parallel process distribution is done carefully.

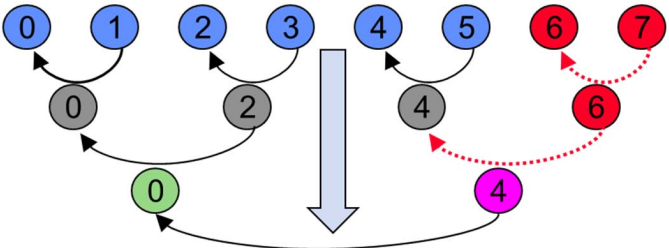

**Figure 6: A conceptual diagram of hierarchical binary aggregation of a distributed global array for the agg() function**. **The arrow denotes the aggregation direction.**

### III. PERFORMANCE BENCHMARKS

The performance of the new file-based messaging kernel using the local filesystems has been compared with that of the current file-based messaging kernel using a central file system. The performance comparison has been made in the three categories: a) point-to-point communication, b) broadcast communication, and c) aggregation communication by running appropriate pMatlab codes running with the MATLAB R2018A release.

In order to measure the performance on the LLSC TX-Green system, a couple of central storage arrays with two different DDN storage hardware arrays running the Lustre parallel filesystem are used. An older SFA 10K DDN storage array [17] is the current production array, which is serving users' home directories. The second storage array is a new SFA 14K DDN storage hardware array [18] which is in testing mode; it will eventually replace the older DDN array for serving users' home directories. It should be noted that the performance measurements are not intended to compare the two different hardware but rather to observe the impact on the file-based message communication performance.

### A. Point-to-Point Communication Performance

The bandwidth and latency measurement of point-to-point communication has been performed by running a pair of parallel processes where one process sends a number of messages with different sizes a total of four times per each message size, and the other process receives the messages. The median value of measured times for each message size is taken in order to avoid any outliers.

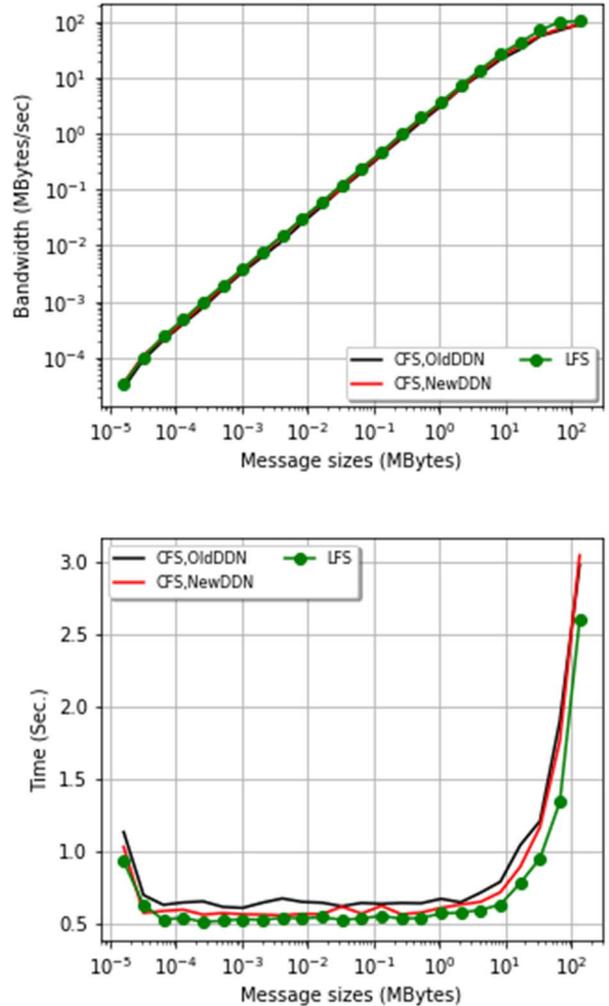

**Figure 7: Bandwidth and execution time as function of the message sizes for a point-to-point message communication within a node. Results are compared with using two different central filesystem (CFS) with different DDN hardware systems (DDN SFA 10K and 14K, respectively) and a local filesystem (LFS).**

Figure 7 shows the performance of the point-to-point communication test, obtained by running both parallel processes on a single node. The x-axis is the message size, while the y-axis is bandwidth (in megabytes per second) and time of messaging. In this case, it does not incur any additional remote file transfer costs since both parallel processes are on the same node. The new DDN hardware for a central filesystem

(CFS) shows slightly better performance at smaller message sizes as compared to the older DDN hardware when looking at the execution time. But the best performance was obtained by using the local filesystem (LFS). The maximum bandwidth was flatten out when the message size is larger than 100 Mbytes. It is interesting to observe that, with the smallest message of 16 bytes, execution times are consistently higher than other messages ranging from 64 bytes to 16 Mbytes for all three filesystems.

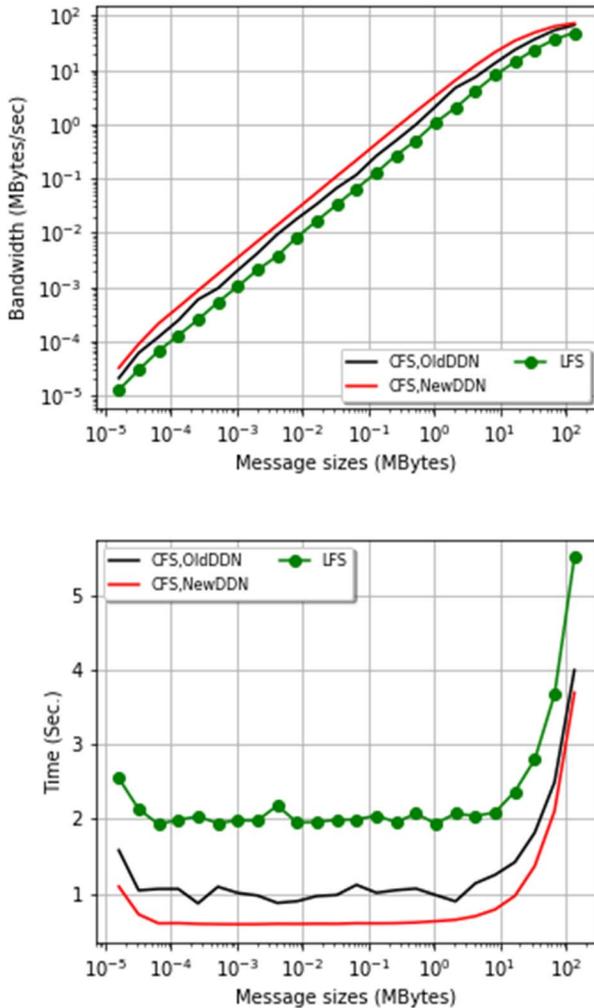

**Figure 8: Bandwidth and execution time as function of the message sizes for a point-to-point message communication between two nodes. Results are compared with using two different central filesystem (CFS) with different DDN hardware and a local filesystem (LFS).**

The next experiment, shown in Figure 8, is point-to-point communication across two compute nodes. This time it is clear that the additional cost for the file transfer between the nodes, when using a local filesystem, causes substantial impact on the bandwidth performance and execution time. This is expected performance degradation due to the additional cost incurred by the file transfer requirement between the nodes. However, the key point is that the new file-based messaging architecture can eliminate significant, if not all, file I/O load on a central filesystem and, therefore, parallel point-to-point communications can avoid any interference from the central filesystem when the central filesystem becomes heavily loaded. In this experiment, the new DDN hardware performs noticeably better than the older DDN hardware. Also, it is noted that the new DDN hardware is not on a production use yet and there is little or no interference from other users, while the older DDN hardware may have significant interference by other users at times since it is in the production mode. This explains why some jittery behavior has been observed with a range of smaller messages on the older DDN hardware.

### B. Broadcast Performance

The broadcast performance was measured by broadcasting a 32-byte message to the pool of various parallel process sizes ranging from 2 to 8192. The time with the current MPI_Bcast() function is faster for smaller number of parallel processes, like $N_p = 2$ and 4 as shown in Figure 9. But the broadcast time grows at much faster rate than that of the node-aware MPI_Bcast() function using two-levels of MPI_Mcast() at smaller numbers of processes, up to 32 processes, which is the maximum number of processes allocated per each compute node in this experiment. When the message is broadcasted beyond a single node, even the new MPI_Bcast() manifests that its broadcast time increase linearly as the number of processes increases although the rate of time increase is smaller than the current MPI_Bcast() implementation. Therefore, the gap between the current and new MPI_Bcast() times becomes greater as the number of processes increases. At $N_p = 1024$, the current MPI_Bcast() function took about 14.3x more time than that of the new MPI_Bcast() function, and at $N_p = 2048$, the current MPI_Bcast() function took about 34x more time.

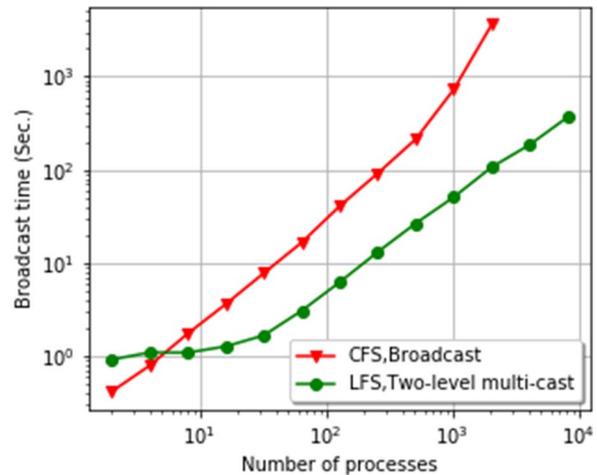

**Figure 9: The time for broadcasting a 32 byte message using a central filesystem (CFS) and local filesystem (LFS) as a function of the number of parallel processes.**

Although the new MPI_Bcast() function has significantly improved performance as compared to the current MPI_Bcast()

function, it still faces a scalability issue if the number of parallel processes increases beyond 100,000 since its broadcast time increases linearly. In order to reduce the broadcast time further, we must address the first level MPI_Mcast() time, which increases linearly as the number of pMatlab processes grows while the second level MPI_Mcast() time remains the same since the number of pMatlab processes per node is capped.

*C. Aggregation Performance*

In order to see the impact of using the new point-to-point communication architecture on the aggregation performance, the time for the aggregation function, agg(), has been measured for a number of globally distributed arrays with various sizes, ranging from 128 Kbytes to 1 Gbytes, as a function of number of parallel processes.

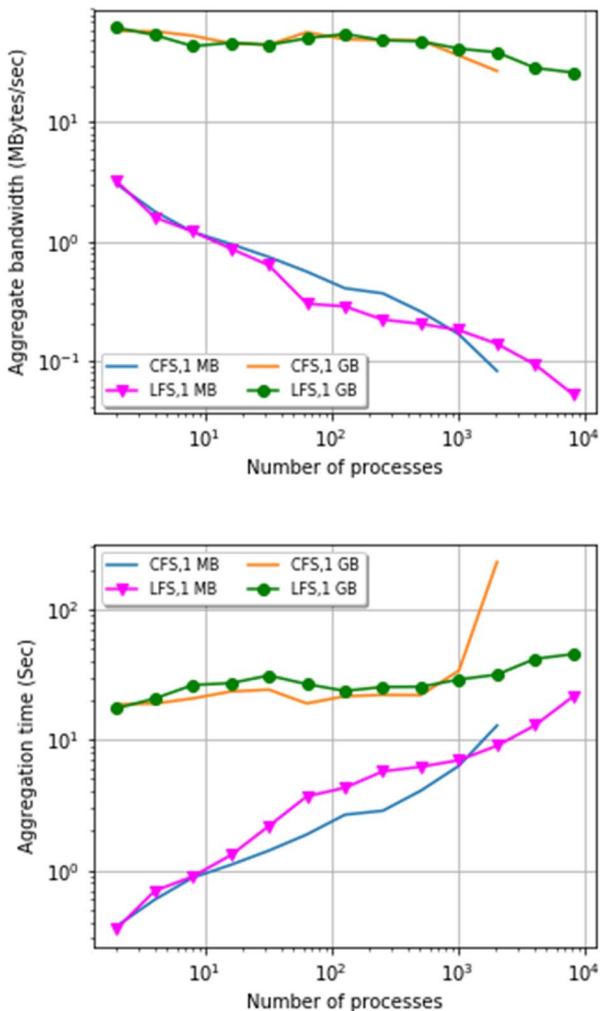

**Figure 10: Aggregation bandwidth and total time for three different globally distributed arrays of 1 Mbytes and 1 Gbytes using a central filesystem (CFS) and local filesystem (LFS) as a function of the number of parallel processes.**

Figure 10 shows the average aggregation bandwidth and total time for the agg() operation on the globally distributed array sizes of 1 Mbytes and 1 Gbytes only. It is interesting to note that the current agg() function performs noticeably better between 16 to 512 parallel processes with the global distributed array of 1 Mbytes size. This is equivalent to gathering the distributed global array on a number of nodes, ranging from 1 to 16 compute nodes. However, for the 1 Gbyte distributed global array, the performance difference between the two is negligible up to 1024 processes. But, it is clear that, beyond $N_p$ = 1024, the agg() performance using the local filesystem (LFS) is outperforming as compared to its performance of using the central filesystem (CFS). This shows that the central filesystem (running the Lustre parallel filesystem) performs best when the number of parallel processes is below 1024. When the number of parallel processes is greater than 1024, the aggregation operation takes significantly longer with the central filesystem, which indicates that the central filesystem is not able to serve the file I/O requests generated by the agg() function. In turn, this interference from the central filesystem degrades the aggregation performance.

SUMMARY

In this paper, we have demonstrated a new file-based message communication architecture which eliminates the use of a central filesystem by using the local filesystem instead. This change avoids any central filesystem loads and resource contention issues and provides better message communication performance for large scale parallel jobs.

We have compared the performance of the new file-based message communications in three categories: A) point-to-point communication, B) broadcast communication and C) aggregation. As we have demonstrated in the performance benchmarks, the overhead costs associated with the additional file transfer requirement with using the local filesystem for file-based message communications is marginal as compared to using the central filesystem. However, its benefit for using the local filesystem for file-based message communications is significant in the broadcast and aggregation performance, especially for a large size parallel jobs.